\documentclass[a4paper,11pt]{article}

\usepackage{nccmath} 
\usepackage{slashed}

\usepackage{pos}

\newcommand{\ep}{\epsilon}

\title{Three-loop four-particle QCD amplitudes}

\author*[a]{Piotr Bargie\l{}a}

\affiliation[a]{Rudolf Peierls Centre for Theoretical Physics, University of Oxford,\\
Clarendon Laboratory, Parks Road, Oxford OX1 3PU, U.K.}

\emailAdd{piotr.bargiela@physics.ox.ac.uk}

\abstract{We present recent advancements in the computation of
  three-loop four-particle helicity amplitudes in full-color massless
  QCD. In this contribution, we focus on the $gg \to \gamma\gamma$
  process. We show how to obtain compact analytic formulae for the
  three-loop scattering amplitude. Our results can be expressed in
  terms of harmonic polylogarithms, which allows for an efficient
  numerical evaluation. The results presented here can be used for improving theoretical predictions relevant for Higgs physics at hadron colliders.}

\FullConference{%
  41st International Conference on High Energy physics - ICHEP2022\\
  6-13 July, 2022\\
  Bologna, Italy
}

\begin{document}
\maketitle

\section{Introduction}

This year, we are celebrating the 10$^{\text{th}}$ anniversary of the
Higgs boson discovery at the Large Hadron Collider
(LHC)~\cite{ATLAS:2012yve,CMS:2012qbp}.  With this achievement,
the particle content of the Standard Model has been confirmed.
Nonetheless, further efforts are still necessary in order to well
establish all properties of fundamental interactions.  Experimentally, it
requires performing high-precision measurements of physical
observables.  Theoretically, this precision needs to be matched by
Standard Model predictions.
At the LHC, most of the real and virtual radiation is due to strong
interactions, described by Quantum Chromodynamics (QCD). Accurate
theoretical predictions require higher-order perturbative
calculations. For virtual corrections, this in turn requires the
knowledge of multi-loop scattering amplitudes. 
Only recently, the three-loop precision has been reached in the full-color QCD
for the following processes: $q\bar{q} \rightarrow
\gamma\gamma$~\cite{Caola:2020dfu}, $q\bar{q} \rightarrow
q\bar{q}$~\cite{Caola:2021rqz}, $gg \rightarrow
\gamma\gamma$~\cite{Bargiela:2021wuy}, $gg \rightarrow
gg$~\cite{Caola:2021izf}, $q\bar{q} \rightarrow
gg$~\cite{Caola:2022dfa} (listed here chronologically).  Here
we will focus on the computation of the amplitude for $gg
\rightarrow \gamma\gamma$~\cite{Bargiela:2021wuy}.

\section{Computation}

\begin{figure}[h]
	\centering
	\includegraphics[width=0.4\textwidth]{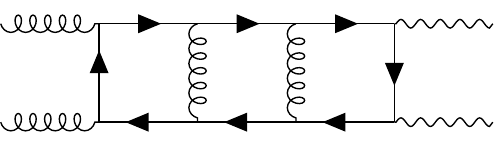}
	\begin{equation*}
	\begin{split}
	& \qquad\qquad\qquad\qquad = \, g_s^6 \, e^2 \, {\color{red} \, n_f^{(V_2)} \, C_f^2 \delta^{a_1,a_2}} \,
	{\color{blue} \int \frac{d^dk_1}{(2\pi)^d} \, \frac{d^dk_2}{(2\pi)^d} \, \frac{d^dk_3}{(2\pi)^d}} \\
	\times& \frac{\color{gray} \text{tr}\left[
		\slashed{\ep}_1 (\slashed{k}_1) \slashed{\ep}_2 (\slashed{k}_1+\slashed{p}_2) \gamma^\mu (\slashed{k}_{13}+\slashed{p}_2) \gamma^\nu (\slashed{k}_2+\slashed{p}_4) \slashed{\ep}_4 (\slashed{k}_2) \slashed{\ep}_3 (\slashed{k}_2-\slashed{p}_3) \gamma_\nu (\slashed{k}_{13}-\slashed{p}_1) \gamma_\mu (\slashed{k}_1-\slashed{p}_1)
		\right]}
	{\color{blue} (k_1)^2 (k_1+p_2)^2 (k_{13}+p_2)^2 (k_2+p_4)^2 (k_2)^2 (k_2-p_3)^2 (k_{13}-p_1)^2 (k_1-p_1)^2 (k_3)^2 (k_{123}-p_{13})^2}
	\end{split}
	\end{equation*}
	\raggedright
	\qquad\quad {\color{red} color} \qquad\qquad\qquad\qquad\quad {\color{gray} tensors} \qquad\qquad\qquad\qquad\qquad\qquad {\color{blue} integrals} \\
	\centering
	\includegraphics[width=0.2\textwidth]{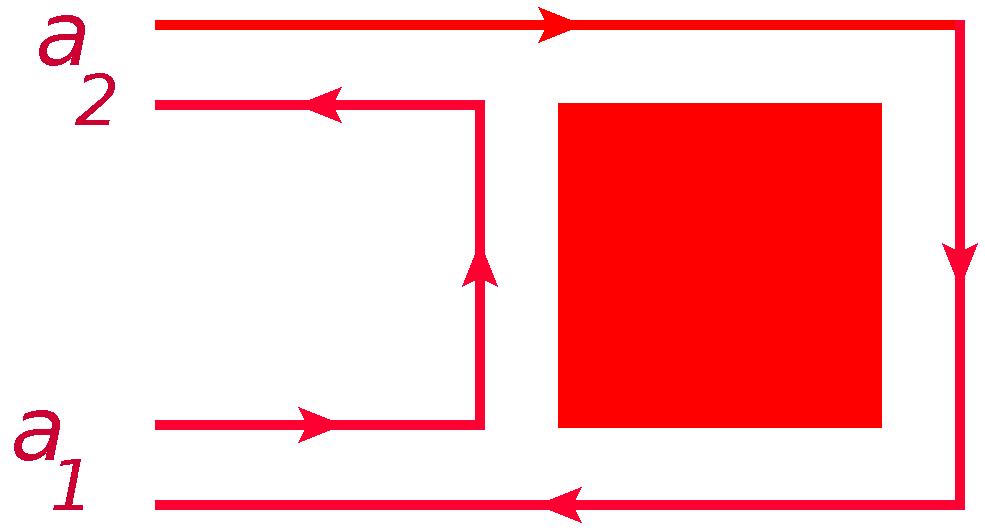}
	\hfill
	\includegraphics[width=0.3\textwidth]{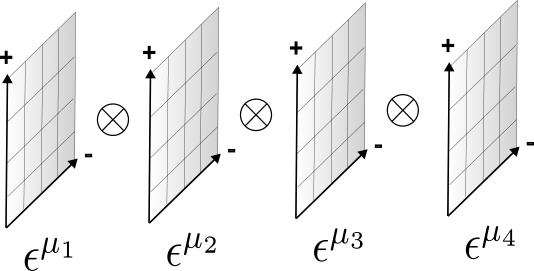}
	\hfill
	\includegraphics[width=0.4\textwidth]{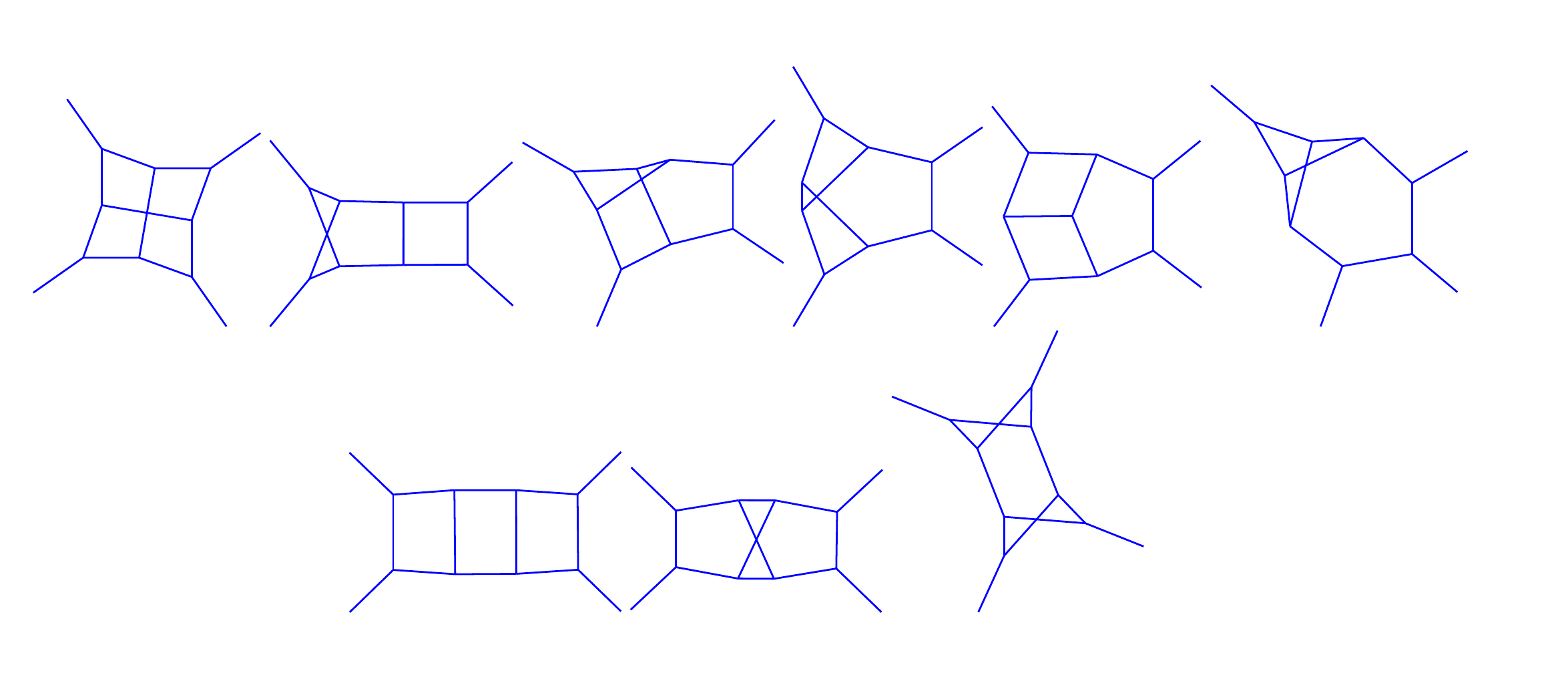}
	\caption{Example Feynman diagram and the three corresponding structures in the amplitude.}
	\label{fig:diag}
\end{figure}

Let us consider the three-loop amplitude for the $g (p_1) + g(p_2) \to
\gamma(-p_3) + \gamma(-p_4)$ process.  At this order, there are 3299
Feynman diagrams, one of which is depicted in Fig.~\ref{fig:diag}.
This number is almost 24 times larger than at the two-loop level, but
still about 15 times smaller than for the $gg \rightarrow gg$ process.
When all Feynman diagrams are summed up to obtain the three-loop
amplitude, the result contains three structures: color, Lorentz
tensors, and Feynman integrals.  The color structure can be extracted
either formally, following $SU(3)$ Lie algebra, or diagrammatically,
using 't Hooft's double line formalism~\cite{tHooft:1973alw}.
For this process, the color algebra does
not pose any challenge. In particular, color factors can be expressed in
terms of quadratic Casimir invariants.
Therefore, from now on 
we will focus on the much more involved tensor and integral structures.

Before computing the required Feynman integrals, we transformed the
integrand into a Lorentz scalar form.  In general, there are
non-trivial Lorentz vector indices arising from external polarization
vectors and $\gamma$ matrices originating from the $\bar{q}\gamma^\mu
q$ vertex structures. Given the large number of Feynman diagrams
involved, these quickly lead to a lot of different Lorentz tensor
structures.  From now on, we will refer to these structures simply as
\textit{tensors}.  If one were to enumerate the independent tensors in
$d$ dimensions for this process at three loops, they would find 138
inequivalent ones. This number can be reduced to 10 by requiring
transversality and by choosing an explicit reference vector for the
external polarisation vectors. On the other hand, the number of
independent helicity states in $d$=4 is $2^4/2$=8, assuming parity
invariance.  Therefore, the number of 10 tensors independent in $d$
dimensions should be possible to further reduce to 8 in four
dimensions. This is indeed possible~\cite{Peraro:2019cjj,Peraro:2020sfm}. In
particular, if one works in the 't Hooft-Veltman scheme, one can
project out two redundant tensors from the four-dimensional subspace
with a standard orthogonalization procedure and only deal with 8 independent
structures. This approach is
loop-universal and it provides a 1-to-1 correspondence between
helicity amplitudes and Lorentz scalar coefficients in the tensor
basis, called \textit{form factors}.

After expressing the amplitude in terms of form factors, one is left
with a large number of three-loop Feynman integrals to evaluate. In
our case, we obtained about $4 \cdot 10^6$ different Feynman
integrals.  In order to compute them in an efficient way, we reduced
them to a minimal basis, and then evaluated only this set. 
They reduced the number of integrals by $\mathcal{O}(20)$.  The most
involved step was to use the Integration-By-Parts (IBP)
relations~\cite{Chetyrkin:1981qh} to project all remaining integrals
onto a basis set of Master Integrals (MIs).  In general, coefficients
in the MI basis are rational functions of the dimension $d$ and
kinematic invariants.  IBP identities follow from the shift invariance
of scalar Feynman integrals, and together with Lorentz invariance
identities generate a linear system~\cite{Laporta:2000dsw}.  In order
to solve such a complicated system, we made use of some modern
mathematical methods.  We exploited $\mathbb{F}_p$ finite-field
arithmetic~\cite{vonManteuffel:2014ixa} to numerically reconstruct
analytic expressions for the rational coefficient functions.  In
addition, we used syzygy-based techniques~\cite{Gluza:2010ws} from
algebraic geometry to constrain the number of redundant integrals
generated by IBP relations.  Moreover, we partial fractioned the
rational coefficient functions in both the dimension $d$ and kinematic
invariants.  It exposed the analytic structure of the amplitude, which
consists of poles and branch cuts in the kinematic invariants coming from
the rational functions and MIs. In
this manner, we obtained a fully analytic decomposition of all $4
\cdot 10^6$ integrals in a basis of 486 MIs.

The analytic expressions for four-particle three-loop massless MIs
have been computed before~\cite{Henn:2020lye}.  Nonetheless, we have
decided to recompute them independently as a check.  The modern
approach to finding analytic expressions for MIs relies on two steps:
constructing a Differential Equation (DE) in the kinematics
invariants, and computing the associated boundary condition. Firstly,
one can find a set of MIs for each of the 9 integral families depicted
in Fig.~\ref{fig:diag} (bottom-right).  Since MIs form a basis, a
derivative of a MI can be further IBP reduced in the same basis of
MIs, thus closing the system of DEs.  With an appropriate choice of MI
basis, the DE has a so-called \textit{canonical form}~\cite{Henn:2013pwa},
and can be easily solved
perturbatively in the dimensional regulator $\ep=(4-d)/2$.
At each $\ep^n$ order, the
result can be written in terms of the well-known Harmonic
Polylogarithms (HPL)~\cite{Gehrmann:2001pz} with only two letters, 0
and 1, in correspondence with poles of the DE.  Secondly, our general
solution to the DE requires fixing the associated boundary
condition. For planar topologies, one could obtain information about
the boundary terms by requiring that the solution does not contain a
particular branch cut in some physical channel, see
e.g. Ref.~\cite{Henn:2014qga}. Since for the complicated
non-planar topologies this method does not work, we relied on the UV
regularity constraint~\cite{Henn:2020lye}.  It follows from an assumption that
Feynman integrals should be regular around each pole of the DE. The DE
can be directly solved in a vicinity of the pole, which generates linear
relations between boundary constants.  In fact, for our DE, all
boundary conditions can be related, order by order in $\ep$, to a single
overall normalization factor. This is in
agreement with Ref.~\cite{Henn:2020lye}. Therefore, in this manner,
we had to compute only one simple overall normalization integral with
direct integration methods in order to reconstruct all the $4 \cdot 10^6$
Feynman integrals in the problem.

\section{Results}

The final expression for the bare scattering amplitude is linear in
HPLs with rational function coefficients, expanded as a series in
powers of $\ep$.  Since the UV and IR behaviour is universal, we can
predict all $\ep$ poles from lower-loop results.  Due to
the Kinoshita-Lee-Nauenberg theorem, these poles cancel exactly
against real emission divergences when combining into the partonic
cross section. Thus, the genuinely new physical information is contained
in the UV-renormalized IR-finite part of the amplitude.  Remarkably,
huge expressions present at intermediate stages of the calculation can
be reduced with aforementioned methods into a compact formula.  For
the simplest helicity configuration, the finite part of the amplitude
reads
\scalebox{0.67}{\parbox{1.0\linewidth}{
		\begin{align*}
		f^{(3,{\rm fin})}_{++++} &= 
		\Delta_1(x)\, n_f^{V_2} C_A^2
		+ \Delta_2(x)\, n_f^{V_2} C_A C_F
		+ \Delta_3(x)\, n_f n_f^{V_2}  C_A
		+ \Delta_4(x)\, (n_f^V)^2 C_A
		+ \Delta_5(x)\, n_f^{V_2} C_F^2 
		+ \Delta_6(x)\, (n_f^V)^2 C_F
		+ \Delta_7(x)\, n_f n_f^{V_2} C_F
		+ \Delta_8(x)\, n_f^2 n_f^{V_2} \nonumber \\
		&\quad + \{(x)\leftrightarrow(1-x)\} \,, \nonumber \\
		\Delta_1(x) &=
		-
		\mfrac{23 L_1 (L_1+2 i \pi )}{9 x^2}+
		\mfrac{32 L_1 (L_1+2 i \pi )-
			46 (L_1+i \pi )}{9x}-
		\mfrac{17}{36} L_0^2-
		\mfrac{19}{36} L_0 L_1+
		\mfrac{1}{9}L_0-2 i \pi L_0
		+\mfrac{1}{288}\pi ^4
		\nonumber \\ & \quad
		-\mfrac{373}{72} \zeta_3
		-\mfrac{185}{72} \pi ^2
		+\mfrac{4519}{324}
		+\mfrac{1}{2}i \pi  \zeta_3
		+\mfrac{11}{144} i \pi ^3
		+\mfrac{157}{12} i \pi 
		+ \mfrac{43}{9} L_0 x
		-\mfrac{7}{9} x^2 \left((L_0-L_1)^2
		+\pi ^2\right)
		\,, \nonumber \\
		\Delta_2(x) &=
		\mfrac{8 L_1 (L_1+2 i \pi )}{3 x^2}
		+\mfrac{16 (L_1+i \pi )
			-{8} L_1 (L_1+2 i \pi )}{3x}
		-\mfrac{1}{3}L_0^2
		+\mfrac{5 }{6}L_0 L_1
		+\mfrac{17}{3}L_0+i \pi  L_0
		-\mfrac{5 }{12}\pi ^2
		-\mfrac{199}{6}
		-{8} i \pi 
		-\mfrac{16}{3} L_0 x
		+\mfrac{4}{3} x^2 \left((L_0-L_1)^2+\pi ^2\right)
		\,, \nonumber \\
		\Delta_3(x) &= 
		\mfrac{L_1 (L_1+2 i \pi )}{18 x^2}
		+\mfrac{2(L_1+i \pi )
			- L_1 (L_1+2 i \pi )}{18x}
		-\mfrac{1}{36}L_0^2
		+\mfrac{1}{36}L_0 L_1
		-\mfrac{1}{9}L_0
		-\mfrac{61 }{36}\zeta_3
		+\mfrac{475}{432} \pi ^2
		-\mfrac{925}{324}
		-\mfrac{1}{72}i \pi ^3
		-\mfrac{175 }{54}i \pi 
		+\mfrac{2}{9} L_0 x
		+\mfrac{1}{36} x^2 \left((L_0-L_1)^2+\pi ^2\right)
		\,, \nonumber \\
		\Delta_4(x) &=
		-\mfrac{5 L_1 (L_1+2 i \pi )}{4 x^2}
		+\mfrac{ L_1 (L_1+2 i \pi )-8 (L_1+i \pi )}{2x}
		+\mfrac{1}{4}L_0^2
		-\mfrac{1}{4}L_0 L_1
		-{2} L_0
		-{6} \zeta_3
		+\mfrac{1}{8}\pi ^2
		-\mfrac{1}{2}
		+{4} L_0 x
		-x^2 \left((L_0-L_1)^2+\pi ^2\right)
		\,, \nonumber \\
		\Delta_5(x) &= 
		-\mfrac{L_1 (L_1+2 i \pi )}{x^2}
		+\mfrac{L_1 (L_1+2 i \pi )-2 (L_1+i \pi )}{x}
		-\mfrac{1}{2}L_0^2
		-i \pi  L_0
		+\mfrac{39}{4}
		+i \pi 
		+{2} L_0 x
		-\mfrac{1}{2} x^2 \left((L_0-L_1)^2+\pi ^2\right)
		\,, \nonumber \\
		\Delta_6(x) &= 
		\mfrac{10 L_1 (L_1+2 i \pi )}{3 x^2}
		+\mfrac{32 (L_1+i \pi )
			-{4} L_1 (L_1+2 i \pi )}{3x}
		-\mfrac{2}{3} L_0^2
		+\mfrac{2}{3} L_0 L_1
		+\mfrac{16}{3} L_0
		+{16} \zeta_3
		-\mfrac{1}{3}\pi ^2
		+\mfrac{4}{3}
		-\mfrac{32}{3} L_0 x
		+\mfrac{8}{3} x^2 \left((L_0-L_1)^2+\pi ^2\right)
		\,, \nonumber \\
		\Delta_7(x) &= 
		\mfrac{5 L_1 (L_1+2 i \pi )}{3 x^2}
		+\mfrac{10 (L_1+i \pi )
			-8 L_1 (L_1+2 i \pi )}{3x}
		+\mfrac{2}{3} L_0^2
		+\mfrac{1}{3}L_0 L_1
		-\mfrac{10}{3}L_0+2 i \pi L_0
		+{4} \zeta_3
		-\mfrac{\pi ^2}{6}
		+{5}
		-{3} i \pi 
		-\mfrac{10 }{3}L_0 x
		+\mfrac{1}{3} x^2 \left((L_0-L_1)^2+\pi ^2\right)
		\,, \nonumber \\
		\Delta_8(x) &= 
		-\mfrac{23 }{216}\pi ^2
		+\mfrac{5 }{27}i \pi 
		\,, \qquad \text{where} \qquad L_0=\ln(x) \,, \quad L_1=\ln(1-x) \,.
		\end{align*}
}}

\noindent
The simplicity of the result partially originates from the lack of
tree-level amplitude for this process.  Moreover, the all-plus
amplitude is just a constant at one-loop order, thus dropping the
highest transcendental weight by two. Still, the complexity of
amplitudes in other helicity configurations does not increase beyond
$\mathcal{O}(10)$ with respect to the expression above.  After
cancelling spurious poles in the kinematic variable $x=-(p_1+p_3)^2/(p_1+p_2)^2$,
these compact expressions can be
efficiently evaluated numerically in $\mathcal{O}(\mu s)$ per each
phase-space point. 

\section{Outlook}

Future directions of investigations are twofold, phenomenological and
formal.  Phenomenologically, the three-loop amplitude for $gg \to
\gamma\gamma$ can be used to compute a fully differential hadronic
cross section. In particular, the three-loop amplitude computed here,
together with two-loop results for $\gamma\gamma$+jet production~\cite{Badger:2021imn}, allow
one to compute theoretical predictions for gluon-induced diphoton
production at the LHC at next-to-next-to-leading order (NNLO).
Although the main diphoton production mechanism at the LHC is through
quark annihilation, the gluon fusion channel is interesting because it
interferes with the Higgs boson amplitude and such an interference can
be used to put bounds on the value of lifetime of the Higgs
boson~\cite{Martin:2013ula,Dixon:2013haa}. On the formal side, the
described methods can be used for other three-loop four-particle
massless processes. Computing three-loop QCD corrections to $\gamma$+jet
production and light-by-light scattering would complete the set of
$2\to 2$ massless scattering amplitudes at this perturbative order.  In
addition, it would be interesting to understand why these processes
require only one independent boundary MI.  The answer to this question
may help us unveil the hidden amplitude structure.

\acknowledgments

PB is thankful to all the authors of three-loop four-particle amplitude publications: Fabrizio Caola, Amlan Chakraborty, Giulio Gambuti, Andreas von Manteuffel, and Lorenzo Tancredi. The research of PB was supported by the ERC Starting Grant 804394 HipQCD. Graphs were drawn with JaxoDraw~\cite{Vermaseren:1994je,Binosi:2003yf}, Qgraf-XML-drawer~\cite{qraf:drawer}, and Inkscape~\cite{Inkscape}.

\bibliographystyle{JHEP}
\bibliography{refs}

\end{document}